# Advancing Waterfall Plots for Cancer Treatment Response Assessment through Adjustment of Incomplete Follow-Up Time

*Zhe Wang[1], Linda Z. Sun[1], Cong Chen[1]*

## Abstract:

Waterfall plots are a key tool in early phase oncology clinical studies for visualizing individual patients' tumor size changes and provide efficacy assessment. However, comparing waterfall plots from ongoing studies with limited follow-up to those from completed studies with long follow-up is challenging due to underestimation of tumor response in ongoing patients. To address this, we propose a novel adjustment method that projects the waterfall plot of an ongoing study to approximate its appearance with sufficient follow-up. Recognizing that waterfall plots are simply rotated survival functions of best tumor size reduction from the baseline (in percentage), we frame the problem in a survival analysis context and adjust weight of each ongoing patients in an interim look Kaplan-Meier curve by leveraging the probability of potential tumor response improvement (i.e., "censoring"). The probability of improvement is quantified through an incomplete multinomial model to estimate the best tumor size change occurrence at each scan time. The adjusted waterfall plots of experimental treatments from ongoing studies are suitable for comparison with historical controls from completed studies, without requiring individual-level data of those controls. A real-data example demonstrates the utility of this method for robust efficacy evaluations.

Keywords: Waterfall plots, Early phase oncology study, Efficacy comparison, Weighted Kaplan-Meier curve.

[1] *Merck & Co., Inc., Rahway, NJ, USA*

# 1. Introduction

In oncology drug development, waterfall plots have emerged as a widely used tool to visualize individual patients' tumor size measurements. These graphical summaries, as demonstrated by early adopters (e.g., Campbell et al., 2007, Socinski et al., 2008, and Kwak et al., 2010), provide an intuitive and reliable method for assessing the efficacy of experimental treatments. A waterfall plot takes the form of a bar chart, where each bar represents a patient's best post-treatment tumor size change (BTSC) from baseline, typically displayed as a percentage. Bars are ordered from left to right based on descending BTSC values, creating the characteristic "waterfall" appearance. Negative values represent tumor reductions, while positive values indicate tumor growth.

Comparing waterfall plots of an experimental treatment from an ongoing study vs. historical controls from completed studies is desirable for informed decision-making to expedite drug development. Huang et al. (2023) proposed a method to visualize and compare waterfall plots from different studies, highlighting the value of cross-trial comparison. However, these comparisons are often confounded by differences in follow-up durations, rendering them not "apple-to-apple." To address this, we propose a direct adjustment to the waterfall plot of an ongoing study, projecting it to approximate the plot that would be observed with long-term follow-up. This approach enables comparisons with historical controls without requiring individual patient-level data from the completed studies, thereby facilitating robust evaluations of treatment efficacy.

During an interim look of an early phase oncology study (i.e., short follow-up), ongoing patients may still achieve further tumor reductions, causing their current BTSC (cBTSC) to underestimate their final BTSC (fBTSC). The fBTSC reflects the best tumor size changes observed after all patients have discontinued treatment and tumor scans, offering a more complete efficacy assessment. The precise modeling of cBTSC and fBTSC is inherently complex, influenced by factors such as tumor type, treatment mechanisms, and patient characteristics. To the best of our knowledge, there is no well-established models to predict tumor size change over time for each individual patient. We tackle this challenge in an alternative way. Instead of predicting the future tumor data for each individual patient, we predict the shape of the fBTSC by leveraging a straightforward relationship between fBTSC and cBTSC at interim.

This relationship is that, over time, the best tumor size change can either improve or maintain its value in the interim analysis (i.e., fBTSC<=cBTSC). Inspired by Sun's paper (2021), which pointed out that a waterfall plot is essentially a transformed survival function of BTSC, we recognized that the situations of "improving" (i.e., fBTSC<cBTSC) and "maintaining" (i.e., fBTSC=cBTSC) can be conceptualized as 'censoring' and 'event', respectively, which are commonly used terminologies in survival analysis. Therefore, the problem boils down to two tasks: 1) to quantify the probability of "event", or equivalently to quantify the probability of "censoring", as these two probabilities add up to 1, and 2) to account for the uncertainty of "event" and "censoring" when constructing the Kaplan-Meier survival curves.

The remainder of this paper is organized as follows. In Section 2, we utilize incomplete multinomial models for interim data and present a Bayesian estimator for the probabilities of 'event' and 'censoring', respectively. Section 3 introduces weighted Kaplan-Meier (KM) curves, with weights accounting for the uncertainty of 'censoring' or 'event'. Through rotating and reversing, weighted KM curves are transformed into waterfall plots, adjusted for ongoing patients' tumor response. Section 4 is an application in a real dataset. Concluding remarks are provided in Section 5.

## 2. Probabilities of "event" and "censoring"

In this section, we set up the problem and establish a statistical model to estimate the probabilities of "event" and "censoring".

### 2.1 Setup

Let us start with variables which are observable from the interim data. For an individual patient i, let $Z_i$ denote the -cBTSC, and $U_i$ denote at which post-treatment scan (e.g., "$U_i$=1" represents the first scan, "$U_i$=2" represents the second scan, and so on.) the cBTSC occurred. For patients actively ongoing in the clinical study, fBTSC can only occur either at the same scan as the cBTSC or at a scan after the interim look. We use $S_i$ to represent the set of scans at which the fBTSC may occur, and it follows that $U_i \in S$. Notably, for patients who already discontinue treatment and stop tumor scanning, the set $S_i$ consists of a single element $U_i$.

To illustrate how these mathematical notations work, we present a toy example involving six patients in total. Please note this example is purely for illustration purposes and is intentionally simplified. Table 1 shows the post-treatment tumor size change (%) for individual patients. Patient #2 and #4 are ongoing patients, while the others already discontinued. An asterisk sign is placed right next to each patient's cBTSC value (i.e., $-Z_i$), which indicates $Z_1 = -30, Z_2 = -10, Z_3 = 0, Z_4 = 25, Z_5 = 35, Z_6 = 90$. Correspondingly, cBTSC's happened at the following scans: $U_1 = 1, U_2 = 3, U_3 = 2, U_4 = 3, U_5 = 3, U_6 = 4$. In this example, ongoing patietns have 3 scans before the interim, and their cBTSC's happened at "scan 3". Therefore, their fBTCS can only happen at scan 3, scan 4 or afterwards (i.e., $S_2 = S_4 = \{3, 4, 5, 6 \ldots \}$).

|  | Scan 1 | Scan 2 | Scan 3 | Scan 4 | Scan 5 |
|---|---|---|---|---|---|
| Patient #1 | **30*** | 35 | discontinue | | |
| Patient #2 | 30 | 20 | **10*** | | |
| Patient #3 | 5 | **0*** | 5 | discontinue | |
| Patient #4 | 0 | -10 | **-25*** | | |
| Patient #5 | 10 | 0 | **-35*** | discontinue | |
| Patient #6 | 0 | -40 | -60 | **-90*** | discontinue |

Table 1 (Toy Example) Patients' post-treatment tumor size change from baseline (%)

Let $\tilde{Z}_i$ be negative fBTSC, and $\tilde{U}_i$ be the scan at which fBTSC will occur. Both $\tilde{Z}_i$ and $\tilde{U}_i$ are observable only if the patient discontinued before interim, and in this case, $\tilde{Z}_i = Z_i$, $\tilde{U}_i = U_i$.

In practice, multiple scans may yield the same tumor size change which equals the greatest tumor reduction. We consider the earliest scan as the "best", so that only one scan, instead of multiple scans, corresponds to the cBTSC or fBTSC.

## 2.2 Incomplete Multinomial Model

The fBTSC is less than or equal to the cBTSC (i.e., $\tilde{Z}_i \geq Z_i$). Moreover, both $\tilde{U}_i$ and $U_i$ belong to the set $S_i$, but they may or may not be the same. Borrowing terminologies from survival analysis, the situation of "$\tilde{Z}_i > Z_i$" is termed as "censoring" and when "$\tilde{Z}_i = Z_i$", it is termed as "event", because the fBTSC value is conclusively observed at interim.

The terms of "event" and "censoring" can also be expressed using the $U_i$ and $\tilde{U}_i$ notations. Then "event" corresponds to $\tilde{U}_i = U_i$, indicating that the patient will not have further improvement in tumor size change so that the scan at which the cBTSC occurred is indeed the one which fBTSC will occur. On the other hand, "censoring" corresponds to $\tilde{U}_i \neq U_i$, meaning fBTSC will occur at a scan after the cBTSC scan.

Consequently, the probabilities of "event" and "censoring" can be denoted as
$p_i = P(\tilde{Z}_i = Z_i) = P(\tilde{U}_i = U_i)$, and
$q_i = P(\tilde{Z}_i > Z_i) = P(\tilde{U}_i \neq U_i)$,
respectively, and $p_i + q_i = 1$.

Predicting $\tilde{Z}_i$ (i.e., negative fBTSC) poses a challenge, requiring a model to predict individual patient tumor size change over time. Recognizing the complexity involved, we adopt an alternative approach: rather than directly predicting the value of fBTSC, we model the scan at which the fBTSC will occur (i.e., $\tilde{U}_i$). This strategy enables us to quantify the uncertainty associated with the situations of 'event' and 'censoring.'

A natural choice for modeling $\tilde{U}_i$ is multinomial models with categories $1, 2, \ldots K$ and category probabilities $\boldsymbol{\theta} = (\theta_1, \ldots \theta_K)$, where $K$ can take the maximum number of scans in a trial and $(\theta_1 + \cdots + \theta_K) = 1$. Then, conditional on interim data, $\tilde{U}_i$ follows the multinomial distribution with categories $S_i$ and category probabilities proportional to $\theta_k$ for $k \in S_i$. It follows that the probabilities of 'event' and 'censoring' can be expressed as

$$p_i = \frac{\theta_{U_i}}{\sum_{k \in S_i} \theta_k}, \quad (1)$$

$$q_i = \frac{\sum_{k \in S_i, k \neq U_i} \theta_k}{\sum_{k \in S_i} \theta_k}, \quad (2)$$

respectively.

A challenge arises as $\tilde{U}_i$ remains unknown for each ongoing patient; we only know that $\tilde{U}_i \in S_i$. Due to the incompleteness of ongoing patients' data, this circumstance leads to an incomplete multinomial model on $S_i$ (see Ahn et al., 2010).

A likelihood function is provided as

$$L(\boldsymbol{\theta}|interim\ data) = \prod_i \sum_{k \in S_i} \theta_k. \qquad (3)$$

Obtaining a maximum likelihood estimator (MLE) by directly optimizing (3) is typically challenging. In the Appendix 1, we introduce a Bayesian estimator of category probabilities as developed by Ahn et al. (2010).

Before jumping to the Section 3, let us consider a trivial aspect of $K$, the number of categories in the multinomial models. Suppose that ongoing patients' cBTSC occurred during the first $J$ scans (i.e, $J = \max(U_i;$ where $i$ refers to each ongoing patient$) = J$), when constructing the incomplete multinomial model, it is convenient to merge the $(J+1)^{th}$ scan and all scans afterwards together, resulting in a total of $J+1$ categories. In other words, we setup $K$ as $J+1$, instead of the maximum number of scans in a study. The interpretation of category $K$ in the multinomial model becomes the combination of $(J+1)^{th}$ scan and all scans afterwards. Such choice of $K$ does not impact calculating probabilities of "event" and "censoring" via (1) and (2), because of the summation structure in the denominator. In the following, without increasing notation complexity, we use scan $J+1$ to refer to the combination of scans at $J+1, J+2\ldots$.

In the toy example mentioned earlier, the total number of categories is set to be K=4 because ongoing patients' cBTSC occurred at the third scan (i.e., $J = \max(U_2, U_4) = 3$). It follows that $S_2 = S_4 = \{3,4\}$), where the number 4 implies the combination of scan 4 and all scans afterwards. Applying (3) in the toy example, the likelihood function becomes $\theta_1 \theta_2 \theta_3 \theta_4 (\theta_3 + \theta_4)^2$, which can be maximized explicitly. The MLE of the category probabilities are $\theta_1 = \theta_2 = \frac{1}{6}$, and $\theta_3 = \theta_4 = \frac{1}{3}$. Consequently, the patients #2 and #4 are both associated with a probability of "event" calculated as $p_2 = p_4 = \frac{\theta_3}{\theta_3 + \theta_4} = \frac{1}{2}$.

## 3. Weighted Kaplan-Meier curves

Understanding the relationship between waterfall plots and survival functions is crucial for our proposed adjustments. As demonstrated in Sun's paper (2021), let $X$ be a generic random variable; then the waterfall plots of $X$ is essentially the survival function of $-X$ after rotation and reverse. To illustrate, consider Plot (1a), which displays the survival curve of $-X$. Rotating this curve counterclockwise by 90 degrees (Plot (1b)) and then reversing it across the horizontal axis (Plot (1c)) transforms it into the waterfall plot of $X$. A simple way to take it is to consider survival curves as stepwise decreasing functions, jumping from 1 to 0 at multiple steps. Then each jump size (from top to bottom) in a survival curve corresponds to each bar height (from left to right) in the waterfall plot. Knowing this relationship, we can make adjustment to survival functions, accounting for the uncertainty of "event" and "censoring," and consequently modify the waterfall plots to accommodate ongoing patients' potential of achieving better tumor responses.

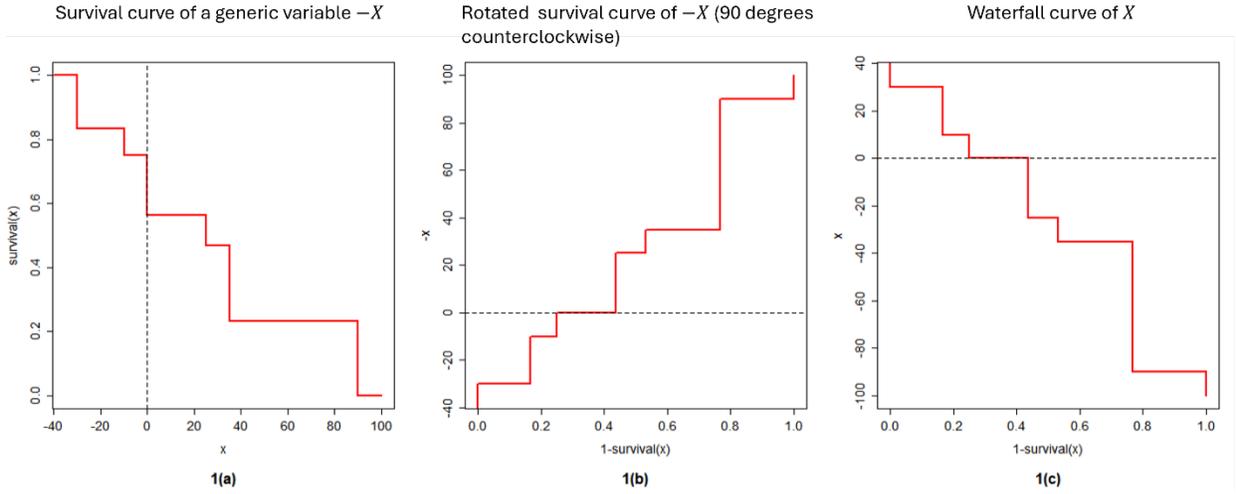

Figure 1 Transformation from the survival curve of $-X$ to the waterfall plot of $X$

Weighted Kaplan-Meier (KM) curves are useful tools for capturing the uncertainty associated with each event. Conventional KM approach places an equal weight to each event, implying a uniform impact on the survival curve. The weighted KM curves, on the other hand, introduce variability by assigning distinct weight to each event. In our context, where the generic variable refers to the -fBTSC (i.e., $\tilde{Z}_i$'s), the weight can take the probability of "event" as estimated in section 2. Let $S_w(z)$ denote the weighted KM curves of -fBTSC, and its formula is given as

$$S_w(z) = \prod_{u \leq z} \left(1 - \frac{dN(u)}{R(u)} p_{i(u)}\right), \quad (4)$$

Both $N(u)$ and $R(u)$ are counting processes, counting the number of $Z_i$'s being smaller than or equal to $u$, and being greater than or equal to $u$, respectively, and $dN(u)$ applies the differentiate operator on $N(u)$. The sub-index of $p_{i(u)}$ refers to the patient associated with $Z_i = u$. Following (4) in the toy example, the weighted KM curve is presented in Figure 1(a), so that Figure 1(c) illustrates the corresponding waterfall plot after adjustment.

Here we present a heuristic interpretation of weighted KM curves. Considering the toy example discussed in Section 2, ongoing patients #2 and #4 have corresponding probabilities of "event" as $p_2 = 0.5$ and $p_4 = 0.5$, and probabilities of "censoring" as $q_2 = 0.5$ and $q_4 = 0.5$. Consequently, four possible scenarios emerge regarding the "event" / "censoring" status of these two patients. As illustrated in Table 2, scenario 1 marks both patients as "events", scenario 2 marks patient #2 as 'event' and patient #5 as 'censoring,' and so forth. The probabilities of each scenario are calculated based on $p_2, q_2, p_4$ and $q_4$, as shown in Table 2. Within each scenario, where the "event" / "censoring" status is certain, a conventional KM curve can be drawn. By averaging these four conventional KM curves, weighted by the scenario probabilities, we precisely obtain what is described in (4). Figure 2 shows the conventional KM curves in each scenario and their weighted average, where the red curve is precisely the survival curve presented in Figure 1(a). This interpretation extends beyond the toy example and remains valid regardless of the number of ongoing patients. A mathematical proof is straightforward, given the nature of the KM curve as a product-limit estimator.

|  | Patient #2 | Patient #4 | Scenario Prob. |
|---|---|---|---|
| Scenario 1 | censoring | censoring | $q_2 q_4 = 0.25$ |
| Scenario 2 | censoring | event | $q_2 p_4 = 0.25$ |
| Scenario 3 | event | censoring | $p_2 q_4 = 0.25$ |
| Scenario 4 | event | event | $p_2 p_4 = 0.25$ |

Table 2 (Toy Example) Scenarios of "event" / "censoring" status of ongoing patients and corresponding scenario probabilities.

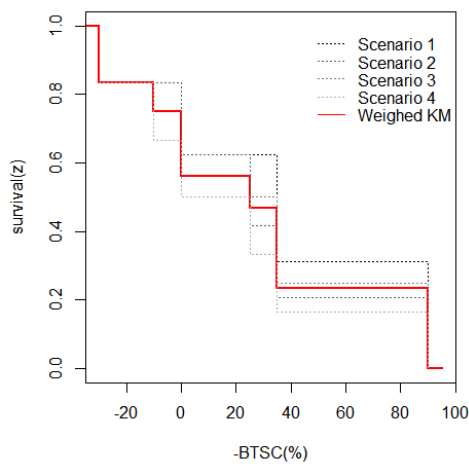

Figure 2 (Toy example) Conventional KM curves in each scenario shown in Table 2 and the weighted KM curve as in formula (4)

The weighted KM curves described in (4) can be computed by R package *survival*, where there is a "*weights*" option for *survfit*. We simply split each patient associated with a non-zero probability of "censoring" (i.e., $0 < q_i$) into two pseudo patients. One pseudo patient has the status as 1, which implies "event" in *survfit*, and the weights as $p_i$, while the other pseudo patient has the status as 0, which implies "censoring" in *survfit*, and the weights as $q_i$. Here we provide an example R code for the toy example.

*toy<- data.frame(*
  *Z=c(-30,-10,-10,0,25,25,35,90),*
  *status=c(1,1,0,1,1,0,1,1),*
  *wt=c(1,.5,.5,1,.5,.5,1,1))*
*survfit(Surv(Z,status)~1, data = toy, weights = wt)*

## 4. Real Data Analysis

In this section, we apply the proposed method to a Phase 2 oncology clinical study to retrospectively assess early efficacy signal and facilitate the comparison against historical controls. By projecting waterfall plot as if follow-up were sufficient, this method could have helped evaluate whether the experimental treatment demonstrated improvement over the control and supported decision-making on the study continuation. For analysis purpose, we selected a specific data cut corresponding to approximately half of the planned enrollment. At this point, 37 patients had been enrolled and had at least one tumor scan, 26 of whom were actively ongoing with tumor scans, while 11 had already discontinued their involvement. Analysis results and related computational considerations are provided in Section 4.1 and 4.2, respectively.

**4.1 Analysis Results**

We assessed the performance of the incomplete multinomial model. In this context, category probabilities $\theta = (\theta_1, \theta_2, \theta_3, \theta_4, \theta_5)$ represent the probabilities that fBTSC occurs at the first, second, third, fourth, or subsequent scans. Employing the framework and estimation techniques outlined in Section 2, we obtain an estimate of category probabilities as $\theta = (0.35, 0.2, 0.25, 0.1, 0.1)$. Notably, these values align closely with the actual frequencies of fBTSC occurrence observed from data, which are $(0.32, 0.22, 0.16, 0.03, 0.27)$, with 12 occurrences at the first scan, 8 at the second scan, 6 at the third scan, 1 at the fourth scan, and 10 at the fifth scan or later, out of the total 37 patients enrolled before cutoff date. As a comparative assessment, simply considering interim cBTSC occurrences does not yield accurate category probabilities. The frequencies of cBTSC occurrences among patients enrolled before the selected data cut are $(0.51, 0.19, 0.24, 0.05, 0)$, while among patients discontinued before cutoff date, they are $(0.63, 0.18, 0.18, 0, 0)$. These two naïve estimations tend to enlarge the probabilities associated with the first scan.

The following Figure 3 presents the outlines of several waterfall plots (i.e., waterfall curves) along with their corresponding 95% point-wise confidence intervals (CIs). The upper and lower bounds are obtained by rotating and reversing the survival curve confidence bounds outputted by *survfit*, following the same transformation detailed in section 3. Figure 3(a) displays the waterfall curve with adjustment on ongoing patients, while Figure 3(b) shows the unadjusted waterfall curve, directly plotting patients' cBTSC values as bar heights. To illustrate how our adjustment aligns with the true fBTSC distribution, both figures include the waterfall plot of actual fBTSC for the 37 enrolled patients as a "ground truth", depicted by the red dotted curve. Evidently, the adjusted waterfall curve is closer to the "ground truth" than the unadjusted curve, and its' CI has a moderate width to effectively cover the "ground truth". In contrast, the CI lower bound of the unadjusted waterfall curve barely covers the "ground truth" curve.

To evaluate the robustness of our method and ensure the observed performance is not due to chance, we generate multiple data replications using all enrolled subjects in this Phase 2 trial by study completion. In each data replication, we shuffled treatment start dates among these patients while preserving each individual's BTSC values and actual time intervals between treatment initiation and tumor scans. The same data cut is applied across data replications, ensuring a consistent sample size (i.e., N=37) while introducing variability in the patietns and scans included for the analysis.

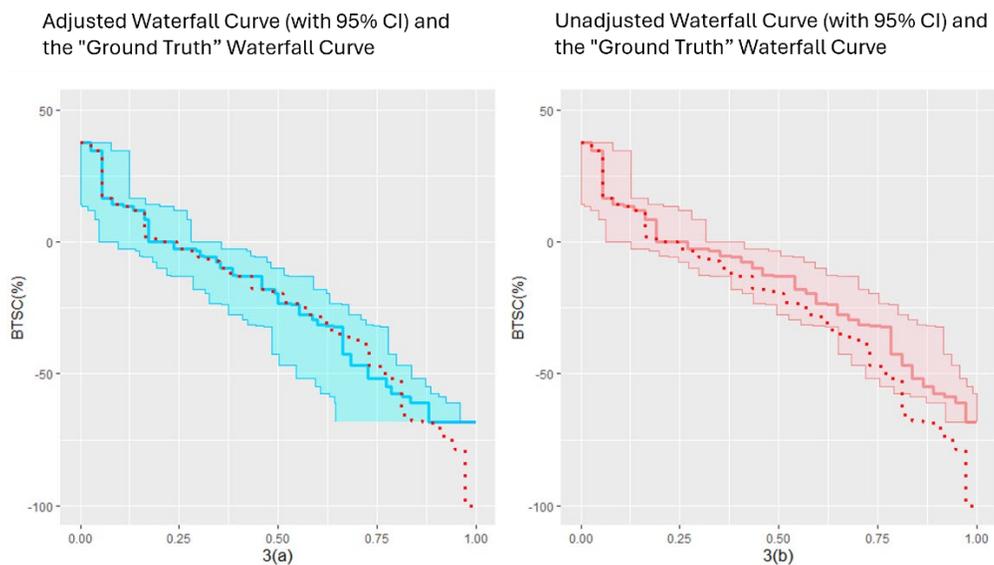

Figure 3 Waterfall curves with and without adjustment on ongoing patients' tumor responses (with 95% CI); Dotted red curve indicates the "ground truth".

Figure 4(a) and Figure 4(b) demonstrate the adjusted and unadjusted waterfall curves across 300 data replications. The black solid curve represents the waterfall curve for all enrolled patients' fBTSC by the study completion (i.e., with long follow-up). If unbiased, the waterfall curves from data replications would be expected to center around the black solid curve, which serves as the ground truth. Notably, the tails of unadjusted curves mostly lie above the solid curve, indicating a tendency to underestimate efficacy. In contrast, the adjusted curves, represented by the blue curves, position the solid curve more centrally, suggesting that the adjustment is consistently effective in addressing this issue across a range of replication scenarios.

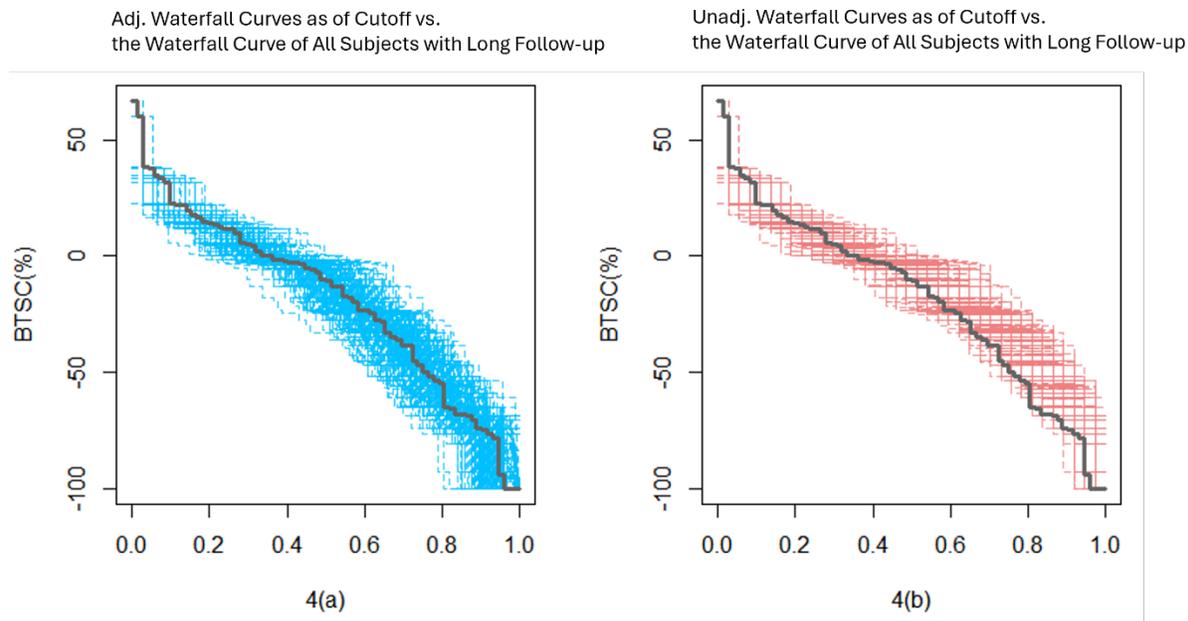

Figure 4 Waterfall curves based on patietns enrolled before data cut vs. all patients enrolled to the study with long follow-up period; Grey solid curve indicates the "ground truth".

### 4.2 Computational Considerations

When implementing method described in Section 2 and Section 3, the first practical consideration arises from the simplicity of the incomplete multinomial model for $\widetilde{U}_i$ outlined in Section 2. Examining the likelihood presented in (3), it is a function of category probabilities $\theta$, conditional on those $S_i$'s observed at interim, which indicates that the model solely takes a part of information (i.e., $S_i$'s) from interim data. While simplicity is a key advantage to our proposed method, an overly simplified model may result in suboptimal performance. To address this, based on our practical experience with various real datasets, we recommend the inclusion of a "filter" for ongoing patients before applying our proposed method. This ensures that ongoing patients are appropriately treated, admitting a patient's potential to achieve deeper tumor response only if one satisfies the filter criteria. In other words, patients who fails to pass the filter are deemed ineligible for achieving further tumor reduction, even if they are ongoing with the study, and their probabilities of "event" are forced to be 1. The details of the filter are described as follows: a) if a patient's most recent two tumor scans remain the same, or b) if a patient has experienced any tumor increase, they do not pass the filter. This refinement aims to strike a balance between simplicity and performance of the model.

The second practical consideration involves the situation where the patients associated with the largest $Z_i$ is ongoing with the study. Section 3 states that each jump size in the weighted KM curves of -fBTSC (i.e., $\widetilde{Z}_i$) corresponds to each bar height in the waterfall plots of fBTSC. However, similar to conventional KM curves, the weighted KM curves may not necessarily decrease to 0 if the patient with the largest $Z_i$ have a non-zero probability of "censoring" ($0 < q_i$). To address this, a straightforward solution is to enforce the weighted

KM curves to drop to 0 for horizontal axis beyond 100, as the best tumor reduction is at most -100%. In other words, -fBTSC cannot exceed 100

## 5. Discussion

The conventional waterfall plots of ongoing studies with limited follow-up, while insightful, cannot take into account the potential for improved tumor response in ongoing patients. Our proposed method addresses this gap by providing a direct adjustment to the waterfall plot, projecting how the plot might look with sufficient follow-up. Our approach avoids the complexities of predicting individual tumor data, offering a practical and efficient solution to account for ongoing patients' potential for deeper tumor reductions as the trial progresses.

The method serves as a forward-looking projection, making it possible to compare waterfall plots from ongoing trials with short follow-up to those from completed trials with mature follow-up. This adjustment effectively levels the playing field, enabling "apple-to-apple" comparisons that are crucial for robust evaluations of treatment efficacy. Of note, our approach focuses solely on adjusting the ongoing trial data, eliminating the need for individual patient-level data from completed studies, which is often unavailable or restricted.

## Appendix: Bayesian Estimator for Category Probabilities

In the method proposed by Ahn et al. (2010), a Gibbs sampler is employed to draw samples from the posterior distribution of $\boldsymbol{\theta}$. A conjunction prior for $\boldsymbol{\theta}$ is Dirichlet distribution $Dir(\boldsymbol{\alpha})$. With $\boldsymbol{\alpha} = (1,..1)$, the prior is flat. The Algorithm 1 describes the Gibbs sampling approach, where posterior samples are denoted as $\boldsymbol{\theta}^{(1)}, \boldsymbol{\theta}^{(2)}, ....$

| | |
|---|---|
| Step 1 | Initialize $\boldsymbol{\theta}^{(0)} = (\theta_1^{(0)}, ... \theta_K^{(0)})$, and ite = 0 |
| Step 2 | For each ongoing patient i, sample $\widetilde{U}_i^{(ite)}$ from a multinomial distribution with categories $S_i$ and category probability proportional to $\theta_k^{(ite)}$ for $k \in S_i$. For discontinued patients, $\widetilde{U}_i^{(ite)}$ is setup as $U_i$. |
| Step 3 | Let the "complete data" be $Y^{(ite)} = (\sum_i I_{[\widetilde{U}_i^{(ite)}=1]}, ..., \sum_i I_{[\widetilde{U}_i^{(ite)}=K]})$, where $I_{[\cdot]}$ is the indicator function. |
| Step 4 | Sample $\boldsymbol{\theta}^{(ite+1)} \sim Dir(\boldsymbol{\alpha} + Y^{(ite)})$ |
| Step 5 | Let ite=ite+1, and repeat Step 2 – Step 4 |

Algorithm 1 Gibbs sampling for posterior distribution of $\boldsymbol{\theta} = (\theta_1, ... \theta_K)$